\documentclass[prd,aps,preprint,showpacs,nofootinbib,superscriptaddress]{revtex4}
\usepackage{epsfig}
\usepackage{amsmath}

\begin{document}


\title{Semi-inclusive Deep Inelastic Scattering at small $x$}
\author{Cyrille Marquet}
\email{cyrille@phys.columbia.edu}
\affiliation{Institut de Physique Th{\'e}orique, CEA/Saclay, 91191 Gif-sur-Yvette cedex, France}
\affiliation{Department of Physics, Columbia University, New York, NY 10027, USA}
\author{Bo-Wen Xiao}
\email{bxiao@lbl.gov}
\affiliation{Nuclear Science Division,
Lawrence Berkeley National Laboratory, Berkeley, CA
94720}
\author{Feng Yuan}
\email{fyuan@lbl.gov} \affiliation{Nuclear Science Division,
Lawrence Berkeley National Laboratory, Berkeley, CA
94720}\affiliation{RIKEN BNL Research Center, Building 510A,
Brookhaven National Laboratory, Upton, NY 11973}

\begin{abstract}
We study the semi-inclusive hadron production in deep inelastic
scattering at small $x$. A transverse momentum dependent
factorization is found consistent with the results calculated in
the small-$x$ approaches, such as the color-dipole framework and
the color glass condensate, in the appropriate kinematic region at
the lowest order. The transverse momentum dependent quark
distribution can be studied in this process as a probe for the
small-$x$ saturation physics. Especially, the ratio of quark
distributions as a function of transverse momentum at different
$x$ demonstrates strong dependence on the saturation scale. The
$Q^2$ dependence of the same ratio is also studied by applying the
Collins-Soper-Sterman resummation method.
\end{abstract}

\maketitle

\newcommand{\be}{\begin{equation}}
\newcommand{\ee}{\end{equation}}
\newcommand{\ben}{\[}
\newcommand{\een}{\]}
\newcommand{\beqn}{\begin{eqnarray}}
\newcommand{\eeqn}{\end{eqnarray}}
\newcommand{\Tr}{{\rm Tr} }

There have been compelling theoretical arguments and experimental
evidence that saturation~\cite{Gribov:1984tu,Mueller:1985wy} plays a very important role in
high-energy hadronic scattering processes, and an effective theory called Color-Glass-Condensate emerges to describe the relevant physics~\cite{McLerran:1993ni,{Iancu:2003xm}}.
In particular, the parton distributions at small-$x$ ($x$ is the longitudinal
momentum fraction of the hadron carried by the parton) and/or of large nucleus
can be calculated from this effective theory, and they all demonstrate
a saturation behavior. The rapidity ($Y=\ln1/x$) evolution of these
distributions are controlled by a nonlinear JIMWLK
equation~\cite{Kovchegov:1999yj,Balitsky:1995ub,JalilianMarian:1997jx},
which has been thoroughly studied in the last decade.
By employing the saturation physics, the deep inelastic
scattering (DIS) structure function measured by the HERA
experiments can be very well described~\cite{GolecBiernat:1998js,Bartels:2002cj,Iancu:2003ge}, as well as the diffractive structure functions~\cite{GolecBiernat:1999qd,Forshaw:2004xd,GolecBiernat:2006ba,Marquet:2007nf} and vector-meson production~\cite{Kowalski:2003hm,Forshaw:2003ki,Marquet:2007qa}.
Forward hadron suppression in $d+Au$ collisions at RHIC experiments also
indicates the importance of saturation in the small-$x$
region~\cite{Kharzeev:2004yx,Dumitru:2005kb,Boer:2007ug,JalilianMarian:2005jf}.
All these successes have encouraged rapid developments in small-$x$ physics
in the last few years~\cite{cgcreviews}.

One of the key predictions of this effective theory is the transverse momentum
dependence of the parton distributions in big nucleus at small-$x$, especially the gluon
distribution~\cite{McLerran:1993ni,McLerran:1998nk,{Mueller:1999wm}}. In the inclusive
DIS process, the gluon distribution is convoluted into a dipole
cross section, which only provides an indirect probe. In this paper, we argue that the transverse momentum
dependent parton distributions can be directly probed
in semi-inclusive processes, for example, in the semi-inclusive hadron production in DIS (SIDIS)~\cite{qiu}. In this process, there are separate momentum scales: the momentum transfered by the virtual photon
squared $Q^2$ and the transverse momentum of the observed hadron in the final state $p_\perp.$
Because of the additional hard momentum scale $Q^2$, the final state hadron transverse momentum
can be directly related to that of the parton distribution
in the nucleon/nucleus when $Q^2$ is much larger than $p_\perp^2$. The relevant QCD factorization
theorem~\cite{Collins:1981uk,Collins:1984kg,{Ji:2004wu}} has
been rigorously studied for the leading-power contribution to the differential cross
section. In the following calculations,
we will extend this factorization argument to the case that involves
saturation physics, and we argue that the transverse momentum
dependent factorization formula is still valid in the so-called
geometric scaling regime~\cite{Stasto:2000er,Iancu:2002tr,Marquet:2006jb,Gelis:2006bs}, when $Q^2$ is much larger than the saturation scale $Q_s^2,$ but saturation effects are still important. As an example, we will demonstrate this factorization for the semi-inclusive DIS process at small-$x$.

\begin{figure}[t]
\begin{center}
\includegraphics[height=4.0cm,angle=0]{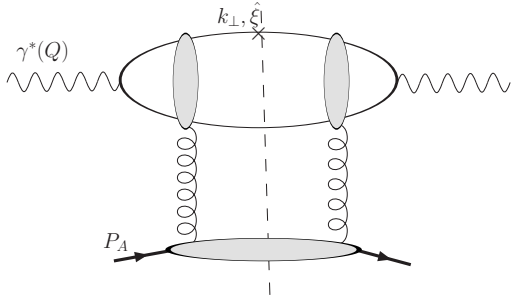}
\vspace*{-0.2cm} \epsfysize=4.2in
\end{center}
\caption{\it Semi-inclusive DIS at small-x, where the cross represents
the quark fragmenting into the final-state hadron. The quark carries the momentum
fraction $\hat\xi$ of the virtual photon, and transverse momentum $k_\perp$.}
\end{figure}

In the semi-inclusive DIS process,
\begin{equation}
e+p(A)\to e'+h+X \ ,
\end{equation}
we observe the final state hadron with
characteristic kinematic variables, such as the longitudinal momentum
fraction $z_h$ of the virtual photon and transverse momentum $p_\perp$.
The usual DIS kinematics variables are defined as $Q^2=-q.q$, $x_B=Q^2/2P_A\cdot q$,
$y=q\cdot P_A/\ell \cdot P_A$, and $z_h=P_h\cdot P_A/q\cdot P_A$,
where $P_h$, $\ell$, $P_A$ and $q$
are momenta for the final state hadron, incoming lepton and nucleon (nucleus), and the exchanged virtual photon, respectively. The transverse momentum $p_\perp$
is usually defined in the center of mass frame of the virtual photon and the incoming hadron.
In Fig.~1, we show the schematic diagram for this
process in the dipole framework at small-$x$~\cite{Mueller:1999wm}, where
the virtual photon splits into a quark-antiquark dipole, then scatters off
the nucleon/nucleus target before the quark (antiquark) fragments into a final state hadron. In the current fragmentation region (forward direction of the virtual photon), the quark-fragmentation contribution will dominate the cross section.

The differential cross section for the above process can be calculated
in the dipole formalism~\cite{Mueller:1999wm} or in the classical Yang-Mills
effective theory approach~\cite{McLerran:1998nk}, and we readily have
\begin{eqnarray}
\frac{d\sigma(ep\to e'hX)}{d{\cal P}}
&=&\frac{\alpha_{em}^2N_c}{2\pi^3x_BQ^2}\sum_f e_f^2\int_{z_h}\frac{dz}{z}\frac{D(z)}{z^2}
\int d^2bd^2q_\perp F(q_\perp,x_B)\times {\cal H}(\hat\xi,k_\perp)\ ,
\end{eqnarray}
where $D(z)$ is the quark fragmentation function into the final
state hadron, $F(q_\perp,x_B)$ the unintegrated gluon distribution
defined below, $\hat\xi=z_h/z$, and the fragmenting quark's
transverse momentum is $k_\perp=p_\perp/z$.\footnote{We could add transverse dependence to the fragmentation function $D(z)$ and write it as $D(z, p_{1\perp})$. In this case, ${\cal H}(\hat\xi,k_\perp)$ should also have $p_{1\perp}$ dependence and becomes ${\cal H}(\hat\xi,\frac{p_\perp-p_{1\perp}}{z})$ accordingly. Here we approximately neglect the transverse dependence of the fragmentation functions. The transverse dependence may change the numeric results, however, it will not affect the following factorization discussion once the above change of variables is made.} The variable $b$ here
is a suppressed variable in $F(q_\perp,x_B)$ which is defined as
the impact parameter with respect to the center of the nucleus. If one
assumes that the nucleus is cylinder like and nucleons are
uniformly distributed inside, one can easily see that the $b$
dependence is trivial and evaluate $\int d^2b$ which yields $\pi
R^2$ with $R$ being the effective radius of the nucleus. The phase
space factor $d{\cal P}$ is defined as $d{\cal
P}=dx_BdQ^2dz_hdp_\perp^2$, and ${\cal H}$ reads as
\begin{eqnarray}
{\cal H}(\hat \xi,k_\perp)&=&\left(1-y+\frac{y^2}{2}\right)(\hat\xi^2+(1-\hat\xi)^2)\left| \frac{k_\perp}{k_\perp^2+\epsilon_f^2}-\frac{k_\perp-q_\perp}{
(k_\perp-q_\perp)^2+\epsilon_f^2}\right|^2\nonumber\\
&&+(1-y)4\hat\xi^2(1-\hat\xi)^2Q^2\left(\frac{1}{k_\perp^2+\epsilon_f^2}-\frac{1}{
(k_\perp-q_\perp)^2+\epsilon_f^2}\right)^2 \ ,
\end{eqnarray}
where $\epsilon_f^2=\hat \xi(1-\hat \xi)Q^2$.
We have also taken the massless-quark limit in the above formula for simplicity,
and the first term is the contribution from transversely polarized photons while the second one
corresponds to longitudinally polarized photons. The unintegrated gluon distribution is defined through the Fourier transform of the dipole cross section:
\begin{equation}
F(q_\perp,x)=\int\frac{d^2r}{(2\pi)^2}e^{-iq_\perp\cdot r} \left(1-T_{q\bar q}(r,x)\right) \ ,
\end{equation}
where $T_{q\bar q}$ is the scattering amplitude, and is
characterized by the saturation scale $Q_s^2$ which depends on
$x$. This unintegrated gluon distribution contains the saturation
physics, which diagrammatically represents the multiple scattering
of the quark-antiquark dipole on nucleon/nucleus target. When
integrating over transverse momentum $p_\perp$ and the
fragmentation function using $\sum_h \int d z D_{q\to h}(z)=1$,
the above formula will reproduce the total DIS cross section in
$ep(A)\to eX$ \footnote{In the $k_t$-factorization at small-$x$,
the gluon momentum fraction $x$ differs from $x_B$ because of the
kinematic constraints~\cite{Kwiecinski:1997ee}. In the leading
logarithmic ($\ln1/x$) approximation, these two are consistent.}.

In this paper, we are interested in the factorization property of the
above differential cross section in the kinematic region where $Q^2$ is much larger
than the final-state hadron transverse momentum $p_\perp^2$. In the current
fragmentation region, $z_h$ is of order 1. Therefore the quark transverse
momentum $k_\perp$ is of the same order as $p_\perp$. Furthermore, we assume
that $Q^2$ is also much larger than the saturation scale $Q_s^2$ which sets the transverse
momentum $q_\perp$ of the unintegrated gluon distribution. Under these limits,
we will be able to study the transverse-momentum-dependent factorization, where we can separate the
transverse momentum dependence of the final-state hadron into the incoming
quark distribution and fragmentation function and/or soft factor~\cite{Collins:1981uk,{Ji:2004wu}}.
An important advantage to utilize the above limits is that we can
apply the power counting to analyze the leading power contribution, and
neglect the higher order corrections in terms of $p_\perp^2/Q^2$ where $p_\perp$ stands
for the typical transverse momentum ($p_\perp\sim k_\perp\sim q_\perp$).

Moreover, we notice that the integral of Eq.~(2) is dominated by the end point contribution
of $\hat \xi\sim 1$ where $\epsilon_f^2$ is in order of $k_\perp^2$~\cite{Mueller:1999wm}.
In order to extract the leading power term from
this equation, we can introduce a delta function in Eq.~(2): $\int d\xi \delta (\xi-1/(1+\Lambda^2/\epsilon_f^2))$
\footnote{If we replace the gluon momentum fraction $x_B$ by $x=x_B/\xi$ in Eq.(2), we will reproduce the $k_t$-factorization formula~\cite{Kwiecinski:1997ee} with this delta function.},
where $\Lambda^2=(1-\hat\xi)k_\perp^2+\hat\xi (k_\perp-q_\perp)^2$, and integrate
out $\hat \xi$ first.
This delta function can be further expanded in the limit of $p_\perp^2\ll Q^2$,
\begin{eqnarray}
&&\delta (\xi-\frac{1}{1+\frac{\Lambda^2}{\epsilon_f^2}})=
\frac{1-\hat \xi}{\xi}\delta\left((1-\xi)(1-\hat \xi)-\frac{\xi}{\hat\xi}
\frac{\Lambda^2}{Q^2}\right)\nonumber\\
&&~~~\to \frac{1-\hat \xi}{\xi}\left(\frac{\delta(1-\hat \xi)}{(1-\xi)}+\frac{\delta(1-\xi)}{(1-\hat \xi)}\right) \ ,
\end{eqnarray}
where a logarithmic term in the above expansion is power suppressed and has been neglected.
The contribution from the second term is also power suppressed.
To see this more clearly, we can substitute $\epsilon_f^2=\xi\Lambda^2/(1-\xi)$ into Eq.~(3),
and the hard coefficient ${\cal H}$ will have an overall factor $(1-\xi)^2$. Combining this
with the delta function expansion, we will find that the second term is the above expansion is
power suppressed relative to the first one.
Applying the delta function expansion in Eqs.~(2) and (3), we will obtain the leading contribution
to the differential cross section in the limit of $p_\perp\ll Q$,
\begin{eqnarray}
\frac{d\sigma(ep\to e'hX)}{d{\cal P}}|_{p_\perp\ll Q}
&=&\frac{\alpha_{em}^2N_c}{2\pi^3Q^4}\sum_f e_f^2\left(1-y+\frac{y^2}{2}\right)
\frac{D(z_h)}{z_h^2}\int\frac{d\xi}{x_B}\nonumber\\
&&~\times
\int d^2bd^2q_\perp F(q_\perp,x_B) A(q_\perp,k_\perp)\ ,
\end{eqnarray}
where
\begin{equation}
A(q_\perp,k_\perp)=\left|\frac{k_\perp|k_\perp-q_\perp|}{(1-\xi)k_\perp^2+\xi(k_\perp-q_\perp)^2}-
\frac{k_\perp-q_\perp}{|k_\perp-q_\perp|}\right|^2 \ .
\end{equation}
We noticed that the longitudinal photon contribution is power suppressed and has been dropped.

On the other hand, a transverse momentum dependent factorization can also be used to describe
the SIDIS process when the hard scale ($Q^2$) is much larger
than the transverse momentum scale $p_\perp^2$.
To leading power of $p_\perp^2/Q^2$, for example,
we will have following factorization formula for the differential cross section
for the semi-inclusive DIS~\cite{Collins:1981uk,{Collins:1984kg},Ji:2004wu},
\begin{eqnarray}
\frac{d\sigma(ep\to e'hX)}{d{\cal P}}
&=&\frac{4\pi\alpha_{em}^2}{Q^2}\left(1-y+\frac{y^2}{2}\right)\int d^2k_{\perp}
d^2p_{1\perp}d^2\lambda_\perp q(x_B,k_\perp;x_B\zeta) D(z_h,p_{1\perp};\hat\zeta/z_h) \nonumber\\
&&\times
S(\lambda_\perp;\rho)H(Q^2,x_B,z_h;\rho)\delta^{(2)}(z_hk_\perp+p_{1\perp}+\lambda_\perp-p_\perp)\
, \label{scros}
\end{eqnarray}
where $q(x_B,k_\perp)$, $D(z_h,p_{1\perp})$, $S(\lambda_\perp)$,
and $H$  are the transverse-momentum-dependent (TMD) quark
distribution, fragmentation function, soft factor, and hard
factor, respectively. We emphasize that the above factorization is
valid in the leading power of $p_\perp^2/Q^2$, and all power
corrections have been neglected.\footnote{The contribution from
the TMD gluon distribution is power suppressed in term of
$p_\perp/Q$ relative to that from the TMD quark distribution in
Eq.~(\ref{scros})\cite{Ji:2004wu} and has been neglected in this
equation.} The energy dependent parameter $\zeta$, $\hat \zeta$
and $\rho$ have been introduced to regulate the light-cone
divergences in the associated functions. In a special frame, we
can simplify them as $x_B^2\zeta^2=\hat\zeta^2/z_h^2=\rho
Q^2$~\cite{Ji:2004wu}. The transverse momentum resummation can be
performed by studying the evolution equation in terms of these
variables. We notice that, since the TMD quark distribution starts
with nontrivial leading order expansion, the TMD fragmentation and
soft factor in Eq.~(\ref{scros}) are trivial at this order:
$D(z_h,p_{1\perp})=D(z_h)\delta^{(2)}(p_{1\perp})$,
$S(\lambda_\perp) =\delta^{(2)}(\lambda_\perp)$. However, at
higher order, for example, the gluon radiation contribution to the
SIDIS process in Fig.~1, we need to take into account nontrivial
expansion of the fragmentation function and the soft factor up to
$\alpha_s$ order\cite{Ji:2004wu}.

The above factorization formalism was studied without considering the small-$x$
resummation effects~\cite{Collins:1981uk,Collins:1984kg,{Ji:2004wu}}.
Here, we assume that the factorization argument can still
hold when the hard momentum scale $Q^2$ is much larger than the saturation
scale $Q_s^2$ and we can use the power counting method to study the leading
contribution in this process. On the other hand, if $Q_s^2$ is the same order
as $Q^2$ (or even larger), the power counting used to argue the TMD factorization
is no longer valid, and we will not have a TMD factorization. \
Similar studies for the heavy quark-antiquark production in $pA$ ($AA$) collisions
have also been discussed in~\cite{Gelis:2003vh}.

\begin{figure}[t]
\begin{center}
\includegraphics[height=8.0cm,angle=0]{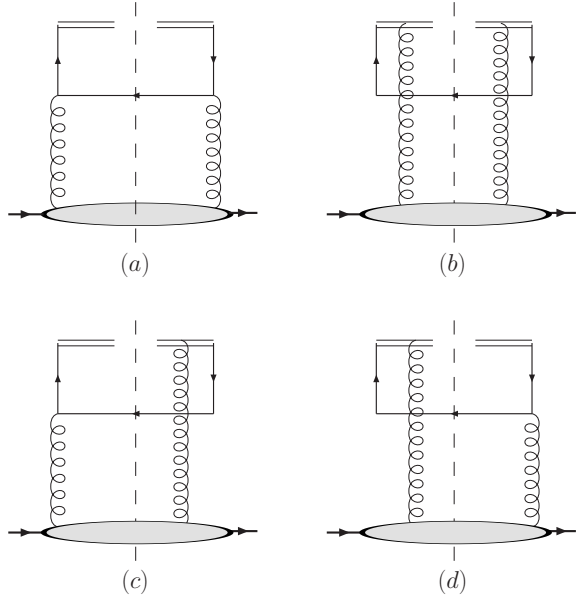}
\vspace*{-0.2cm} \epsfysize=4.2in
\end{center}
\caption{\it Transverse-momentum-dependent quark distribution calculated from small-x gluon splitting.
The double line represents the gauge link contribution from the TMD quark distribution definition.}
\end{figure}

As an important check, in the following we will compare the prediction from
the TMD formula Eq.~(8) to the dipole result Eq.~(6) in the same kinematic region,
$Q^2\gg p_\perp^2(Q_s^2)$.
To do that, we need to calculate the TMD quark distribution in nucleon/nucleus
at small-$x$. This quark distribution is defined as~\cite{Collins:1981uk}
\begin{equation}
q(x,k_\perp)=\frac{1}{2}\int\frac{d^2\xi_\perp d\xi^-}{(2\pi)^2}
e^{-ixP^+\xi^--ik_\perp\cdot \xi_\perp}\langle P|\bar\Psi(\xi){\cal L}_\xi
\gamma^+{\cal L}_0\Psi(0)|P\rangle\ ,
\end{equation}
where $P$ is the momentum for the hadron, $x$ and $k_\perp$ are
longitudinal momentum fraction of the hadron and transverse momentum carried by
the quark. In the above equation, ${\cal L}$ is the gauge link introduced
to guarantee the gauge invariance of the above definition~\cite{Collins:1981uk,{Ji:2004wu}}.
At this particular order, the gluon splitting
contribution to the TMD quark distribution can be calculated
in the $k_t$-factorization approach at small $x$.
We show the relevant Feynman diagrams in Fig.~2, where (b-d) diagrams come from
the gauge link contributions. These diagrams have to be taken
into account because the gauge field connecting
to the hadron state (nucleon/nucleus) are dominated by the $A^+$
component in the $k_t$-factorization calculations.
Their contributions are important to obtain a consistent and
gauge-invariant result. The derivation is straightforward, and we have
\begin{equation}
q(x,k_\perp)=\frac{N_c}{8\pi^4}\int\frac{dx'}{x^{\prime 2}}
\int d^2bd^2q_\perp F(q_\perp,x') A(q_\perp,k_\perp) \ ,
\end{equation}
where $A$ has been defined in Eq.~(7). This is the quark distribution
calculated in the $k_t$-factorization. In order to compare to the results
we obtain above in the color-dipole formalism, we need to extrapolate
in the leading logarithmic approximation at small-$x$, i.e., replacing
the unintegrated gluon distribution $F(q_\perp,x')$ by $F(q_\perp,x)$ in the
above equation. Following this replacement, we will reproduce
the differential cross section Eq.~(6) calculated in the dipole framework at the leading
order of $p_\perp^2/Q^2$. Therefore, we have demonstrated
that the small-$x$ calculation of the differential cross section for the SIDIS process
is consistent with the TMD factorization at this particular order. At even higher order, we will have
to take into account the contributions from the fragmentation function and soft
factor. At this order, they are trivial: $D(z_h,p_{1\perp})=D(z_h)\delta^{(2)}(p_{1\perp})$
and $S(\lambda_\perp)=\delta^{(2)}(\lambda_\perp)$ where $D(z_h)$ is the
integrated fragmentation function. We further argue that the TMD factorization
will work at higher orders as well, because the power counting is valid when
$Q^2\gg p_\perp^2$($Q_s^2$) as we mentioned above.

In the leading logarithmic approximation at small-$x$, we can further integrate out
$\xi$ in Eq.~(10):
\begin{equation}
xq(x,k_\perp)=\frac{N_c}{4\pi^4}\int d^2bd^2q_\perp F(q_\perp,x)\left(1-\frac{k_\perp\cdot(k_\perp-q_\perp)}
{k_\perp^2-(k_\perp-q_\perp)^2}\ln\frac{k_\perp^2}{(k_\perp-q_\perp)^2}\right) \ ,
\end{equation}
which is consistent with the result calculated before~\cite{McLerran:1998nk}.
A number of interesting features of this quark distribution have
been discussed in the literature~\cite{{McLerran:1998nk},Mueller:1999wm}.
For example, in the small $k_\perp$ limit, the quark distribution saturates: $
xq(x,k_\perp)|_{k_\perp\to 0}\propto {N_c}/{4\pi^4}$; in the large $k_\perp$ limit,
it has power-law behavior $xq(x,k_\perp)\big |_{k_\perp\gg Q_s}\propto {Q_s^2}/{k_\perp^2}$.
These two features will be manifest if we employ the GBW model for the
unintegrated gluon distribution: $F(q_\perp,x)=e^{-q_\perp^2/Q_s^2}/\pi Q_s^2$,
where $Q_s^2$ is parameterized as $Q_s^2=(x/x_0)^\lambda GeV^2$ with $x_0=3\cdot 10^{-4}$
and $\lambda=0.28$~\cite{GolecBiernat:1998js}.
Note that while the large $q_\perp$ behavior of the unintegrated gluon distribution $F(q_\perp,x)$ is
incorrect in the GBW model (it falls exponentially instead of a power law), this bad feature does
not translate to the TMD quark distribution: the convolution with the splitting kernel in Eq. (7) insures
the proper leading-twist behavior. In Fig.~3 (left panel),
we show the ratio of the TMD quark distribution $xq(x,k_\perp)$ relative to that at
$x=10^{-2}$ as a function of $k_\perp$ for $x=10^{-4}$ and $x=10^{-3}$, respectively.
From this figure, we can clearly see that the ratio remains unchanged when $k_\perp$ goes to 0,
whereas the ratio is proportional to the ratio of $Q_s^2$ at different $x$
when $k_\perp$ is large. This clearly demonstrates that the transverse
momentum dependence
provides an important information on the saturation physics.
We have shown that these TMD quark distributions can be studied in
semi-inclusive DIS process.

Furthermore, the transverse momentum dependence is also sensitive to the
QCD dynamics of the small-$x$ evolution. In the above example, we took
the simple parameterization from the GBW model~\cite{GolecBiernat:1998js}.
This result shall be modified by the nonlinear evolution. For example, at large $q_\perp$,
the unintegrated gluon distribution scales as $(q_\perp^2/Q_s^2)^{-\lambda_c}$ where
$\lambda_c$ is the anomalous dimension~\cite{Mueller:2002zm,Iancu:2002tr,Kharzeev:2004yx,{Albacete:2003iq}}.
In the DGLAP domain, we have $\lambda_c=1$ whereas in the BFKL domain it is $\lambda_c=0.5$.
By solving the BK equation, it was found that $\lambda_c=0.63$ for large rapidities
$Y=\ln1/x$~\cite{Iancu:2002tr,Munier:2003vc}. With this modification,
the ratio of the TMD quark distribution at large $k_\perp$ will approach $(Q_s^2)^{\lambda_c}$ instead of $Q_s^2$.

\begin{figure}[t]
\begin{center}
\includegraphics[height=5.0cm,angle=0]{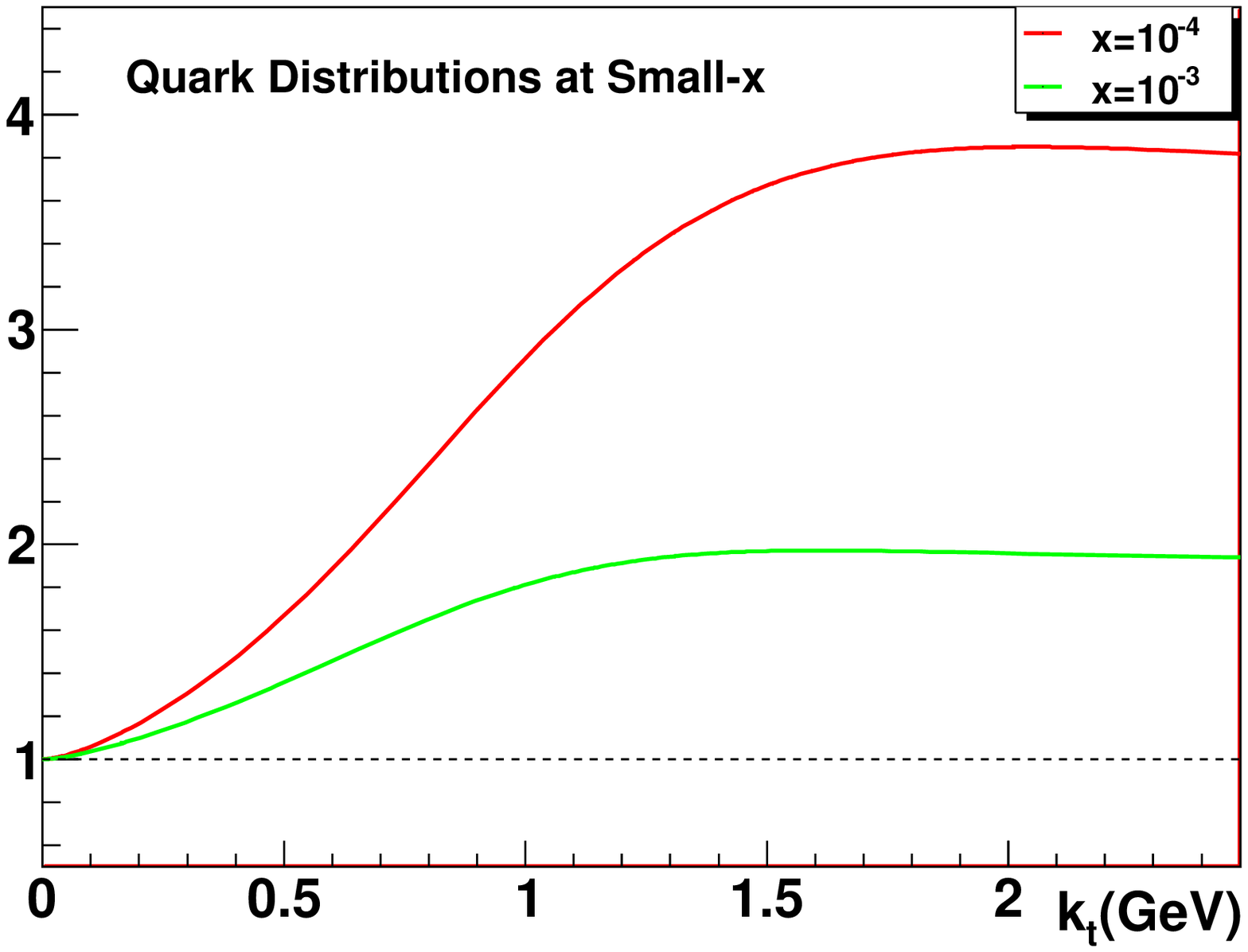}
\includegraphics[height=5.0cm,angle=0]{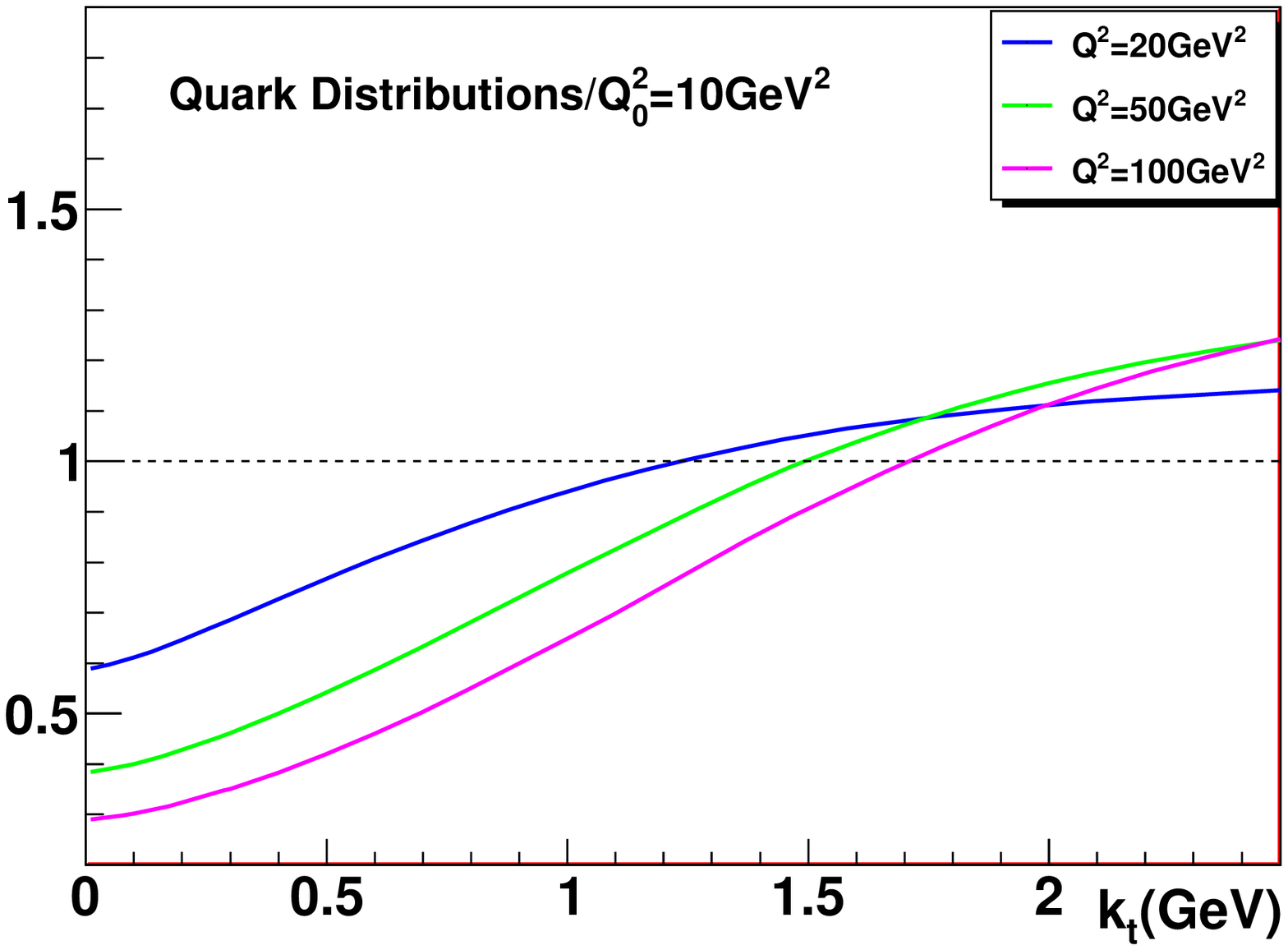}
\vspace*{-0.2cm} \epsfysize=4.2in
\end{center}
\caption{\it The transverse-momentum-dependent quark distributions at small-$x$: (left) at
$x=10^{-4}$ and $10^{-3}$ as ratios relative to that at $x=10^{-2}$ for fixed
$Q^2=Q_0^2=10GeV^2$ where the transverse momentum resummation effect is not important;
(right) at different $Q^2$ relative to that at $Q^2=10GeV^2$ for fixed $x=3\cdot 10^{-4}$ with
$Q_s^2=1GeV^2$ in the GBW model.}
\end{figure}

Another important QCD dynamics effect is the transverse momentum
resummation~\cite{Collins:1984kg}, which will affect the $Q^2$ dependence
of the $k_\perp$ spectrum. In the results we plotted in the left panel of Fig.~3,
this effect was not considered, which correspond to the low $Q^2$ results, for
example, at $Q^2=Q_0^2=10GeV^2$.
This effect can be studied by applying the
Collins-Soper-Sterman resummation method~\cite{Collins:1984kg}.
There have been great applications of this method to various high-energy
processes, in particular, in the semi-inclusive DIS at HERA~\cite{Nadolsky:1999kb}
where important consequences have been observed. To demonstrate this effect
in the TMD quark distribution at small-$x$ we calculated above, we take the double leading logarithmic approximation (DLLA) to solve the evolution equation for the quark distributions. Under this approximation,
we can write down quark distribution at higher $Q^2$ in terms of that at
lower $Q_0^2$~\cite{Idilbi:2004vb,{Boer:2008fr}}\footnote{Here, we approximate the
energy dependent parameter $\zeta$ in the TMD quark distribution by $Q$ assuming
they are of the same order~\cite{Ji:2004wu}.}:
\begin{equation}
q(x,k_\perp;Q^2)=\int\frac{d^2r}{(2\pi)^2}e^{ik_\perp\cdot r}e^{-S(Q^2,Q_0^2,r)}\int
d^2k_{\perp}' e^{-ik_{\perp}'\cdot r}q(x,k_{\perp}';Q_0^2)\ ,
\end{equation}
where the Sudakov form factor at the DLLA is defined by
\begin{equation}
S(Q^2,Q_0^2,r)=\ln\frac{Q^2}{Q_0^2}\left[\frac{\alpha_sC_F}{4\pi}\ln(Q^2Q_0^2 r^4)+c_0 r^2\right] \ ,
\end{equation}
where we have also included a non-perturbative form factor
contribution $c_0 r^2\ln Q^2/Q_0^2$~\cite{Korchemsky:1994is}. This
resummation effect will shift the transverse momentum distribution
to higher end when $Q^2$ increases. As an example, in Fig.~3
(right panel) we show the typical changes for the quark
distributions at $Q^2=20,~50,~100GeV^2$ as compared to the lower
$Q_0^2=10GeV^2$, with the following parameters: fixed coupling
$\alpha_s=0.3$, and $c_0=0.1$ for the non-perturbative input for
the form factor~\cite{Nadolsky:1999kb}. From this plot, we can see
that indeed, the transverse momentum distribution becomes harder
when $Q^2$ is larger. We notice that neglecting the resummation
effects in the left panel of Fig.~3 will introduce additional
theoretical uncertainties in the predictions. However, we expect
that much of the resummation effects will be cancelled out in the
ratios of the quark distributions at the same $Q^2$.

In conclusion, we have studied the semi-inclusive DIS processes at small-$x$,
and found that the quark distribution studied can be used as a probe of saturation physics.
Especially, the ratio of the quark distributions is
crucially depending on the saturation scale. We also studied the quark
distribution at different $Q^2$ values, and found that the resummation effects shift
the distribution to larger $k_\perp$ with larger $Q^2$. An ideal place to study
this physics will be an electron-ion collider in the near future~\cite{Deshpande:2005wd},
where large nucleus target will provide an additional direction to study saturation.
Meanwhile, we notice that the ratios plotted in
Fig.~3 qualitatively agree with the experimental data from HERA~\cite{Adloff:1996dy}.
Of course, in order to compare to these data, we have to take into account
the fragmentation contributions to calculate the differential cross sections.
We also notice that an extension to a study on the gluon transverse momentum
distributions~\cite{Kovchegov:1998bi} will have to consider both
small-$x$ and transverse momentum resummations. The result from
this paper shall provide us confidence to carry out these important
studies.

F.Y. thanks Anna Stasto and George Sterman for interesting conversations a few years back about the topic studied in this paper. C.M. would like to thank the nuclear theory group at LBL for hospitality during the start of this work. We thank Al Mueller, Pavel Nadolsky, Jianwei Qiu, and Raju Venugopalan for stimulating discussions and comments. C.M. is supported by the European Commission under the FP6 program, contract No. MOIF-CT-2006-039860. This work was supported in part by the U.S. Department of Energy under contract DE-AC02-05CH11231. We are grateful to RIKEN, Brookhaven National Laboratory and the U.S. Department of Energy
(contract number DE-AC02-98CH10886) for providing the facilities essential for the completion of this work.


\begin{thebibliography}
\frenchspacing

\bibitem{Gribov:1984tu}
  L.~V.~Gribov, E.~M.~Levin and M.~G.~Ryskin,
  Phys.\ Rept.\  {\bf 100}, 1 (1983).

\bibitem{Mueller:1985wy}
  A.~H.~Mueller and J.~w.~Qiu,
  Nucl.\ Phys.\  B {\bf 268}, 427 (1986).

\bibitem{McLerran:1993ni}
  L.~D.~McLerran and R.~Venugopalan,
  Phys.\ Rev.\  D {\bf 49}, 2233 (1994);
  Phys.\ Rev.\  D {\bf 49}, 3352 (1994);
  Phys.\ Rev.\  D {\bf 50}, 2225 (1994).

\bibitem{Iancu:2003xm}
 see a review, E.~Iancu and R.~Venugopalan,
  arXiv:hep-ph/0303204.

\bibitem{Balitsky:1995ub}
  I.~Balitsky,
  Nucl.\ Phys.\  B {\bf 463}, 99 (1996).

\bibitem{Kovchegov:1999yj}
  Y.~V.~Kovchegov,
  Phys.\ Rev.\  D {\bf 60}, 034008 (1999).

\bibitem{JalilianMarian:1997jx}
  J.~Jalilian-Marian, A.~Kovner, A.~Leonidov and H.~Weigert,
  Nucl.\ Phys.\  B {\bf 504}, 415 (1997);
  Phys.\ Rev.\  D {\bf 59}, 014014 (1999);
  E.~Iancu, A.~Leonidov and L.~D.~McLerran,
  Nucl.\ Phys.\  A {\bf 692}, 583 (2001).

\bibitem{GolecBiernat:1998js}
  K.~J.~Golec-Biernat and M.~Wusthoff,
  Phys.\ Rev.\  D {\bf 59}, 014017 (1999).

\bibitem{Bartels:2002cj}
  J.~Bartels, K.~J.~Golec-Biernat and H.~Kowalski,
  Phys.\ Rev.\  D {\bf 66}, 014001 (2002).

\bibitem{Iancu:2003ge}
  E.~Iancu, K.~Itakura and S.~Munier,
  Phys.\ Lett.\  B {\bf 590}, 199 (2004);
  G.~Soyez,
  Phys.\ Lett.\  B {\bf 655}, 32 (2007).

\bibitem{GolecBiernat:1999qd}
  K.~J.~Golec-Biernat and M.~Wusthoff,
  Phys.\ Rev.\  D {\bf 60}, 114023 (1999).

\bibitem{Forshaw:2004xd}
  J.~R.~Forshaw, R.~Sandapen and G.~Shaw,
  Phys.\ Lett.\  B {\bf 594}, 283 (2004).

\bibitem{GolecBiernat:2006ba}
  K.~J.~Golec-Biernat and S.~Sapeta,
  Phys.\ Rev.\  D {\bf 74}, 054032 (2006).

\bibitem{Marquet:2007nf}
  C.~Marquet,
  Phys.\ Rev.\  D {\bf 76}, 094017 (2007).

\bibitem{Kowalski:2003hm}
  H.~Kowalski and D.~Teaney,
  Phys.\ Rev.\  D {\bf 68}, 114005 (2003);
  H.~Kowalski, L.~Motyka and G.~Watt,
  Phys.\ Rev.\  D {\bf 74}, 074016 (2006).

\bibitem{Forshaw:2003ki}
  J.~R.~Forshaw, R.~Sandapen and G.~Shaw,
  Phys.\ Rev.\  D {\bf 69}, 094013 (2004).

\bibitem{Marquet:2007qa}
  C.~Marquet, R.~B.~Peschanski and G.~Soyez,
  Phys.\ Rev.\  D {\bf 76}, 034011 (2007).

\bibitem{Kharzeev:2004yx}
  D.~Kharzeev, Y.~V.~Kovchegov and K.~Tuchin,
  Phys.\ Lett.\  B {\bf 599}, 23 (2004).

\bibitem{Dumitru:2005kb}
  A.~Dumitru, A.~Hayashigaki and J.~Jalilian-Marian,
  Nucl.\ Phys.\  A {\bf 770}, 57 (2006);
  A.~Dumitru, A.~Hayashigaki and J.~Jalilian-Marian,
  Nucl.\ Phys.\  A {\bf 765}, 464 (2006).

\bibitem{Boer:2007ug}
  D.~Boer, A.~Utermann and E.~Wessels,
  Phys.\ Rev.\  D {\bf 77}, 054014 (2008).

\bibitem{JalilianMarian:2005jf}
  J.~Jalilian-Marian and Y.~V.~Kovchegov,
  Prog.\ Part.\ Nucl.\ Phys.\  {\bf 56}, 104 (2006).

\bibitem{cgcreviews}
  H.~Weigert,
  Prog.\ Part.\ Nucl.\ Phys.\  {\bf 55}, 461 (2005);
  D.~N.~Triantafyllopoulos,
  Acta Phys.\ Polon.\  B {\bf 36}, 3593 (2005);
  G.~Soyez,
  Acta Phys.\ Polon.\  B {\bf 37}, 3477 (2006);
  S.~Munier,
  Phys.\ Rept.\  {\bf 473}, 1 (2009).

\bibitem{McLerran:1998nk}
  L.~D.~McLerran and R.~Venugopalan,
  Phys.\ Rev.\  D {\bf 59}, 094002 (1999);
  R.~Venugopalan,
  Acta Phys.\ Polon.\  B {\bf 30}, 3731 (1999).

\bibitem{Mueller:1999wm}
  A.~H.~Mueller,
  Nucl.\ Phys.\  B {\bf 558}, 285 (1999).

\bibitem{qiu}
The mean transverse momentum in this process
was suggested to study the saturation physics, J.~Qiu,
Electron-Ion Collider Collaboration Meeting, Berkeley, December
11-13, 2008.

\bibitem{Collins:1981uk}
  J.~C.~Collins and D.~E.~Soper,
  Nucl.\ Phys.\  B {\bf 193}, 381 (1981)
  [Erratum-ibid.\  B {\bf 213}, 545 (1983)];
  Nucl.\ Phys.\  B {\bf 194}, 445 (1982).

\bibitem{Collins:1984kg}
  J.~C.~Collins, D.~E.~Soper and G.~Sterman,
  Nucl.\ Phys.\  B {\bf 250}, 199 (1985).

\bibitem{Ji:2004wu}
  X.~Ji, J.~Ma and F.~Yuan,
  Phys.\ Rev.\  D {\bf 71}, 034005 (2005).

\bibitem{Stasto:2000er}
  A.~M.~Stasto, K.~J.~Golec-Biernat and J.~Kwiecinski,
  Phys.\ Rev.\ Lett.\  {\bf 86}, 596 (2001).

\bibitem{Iancu:2002tr}
  E.~Iancu, K.~Itakura and L.~McLerran,
  Nucl.\ Phys.\  A {\bf 708}, 327 (2002).

\bibitem{Marquet:2006jb}
  C.~Marquet and L.~Schoeffel,
  Phys.\ Lett.\  B {\bf 639}, 471 (2006).

\bibitem{Gelis:2006bs}
  F.~Gelis, R.~B.~Peschanski, G.~Soyez and L.~Schoeffel,
  Phys.\ Lett.\  B {\bf 647}, 376 (2007).

\bibitem{Kwiecinski:1997ee}
  J.~Kwiecinski, A.~D.~Martin and A.~M.~Stasto,
  Phys.\ Rev.\  D {\bf 56}, 3991 (1997);
  K.~Golec-Biernat and A.~M.~Stasto,
  arXiv:0905.1321 [hep-ph].

\bibitem{Gelis:2003vh}
  F.~Gelis and R.~Venugopalan,
  Phys.\ Rev.\  D {\bf 69}, 014019 (2004).

\bibitem{Mueller:2002zm}
  A.~H.~Mueller and D.~N.~Triantafyllopoulos,
  Nucl.\ Phys.\  B {\bf 640}, 331 (2002).

\bibitem{Albacete:2003iq}
  J.~L.~Albacete, N.~Armesto, A.~Kovner, C.~A.~Salgado and U.~A.~Wiedemann,
  Phys.\ Rev.\ Lett.\  {\bf 92}, 082001 (2004).

\bibitem{Munier:2003vc}
  S.~Munier and R.~B.~Peschanski,
  Phys.\ Rev.\ Lett.\  {\bf 91}, 232001 (2003).

\bibitem{Nadolsky:1999kb}
  P.~M.~Nadolsky, D.~R.~Stump and C.~P.~Yuan,
  Phys.\ Rev.\  D {\bf 61}, 014003 (2000)
  [Erratum-ibid.\  D {\bf 64}, 059903 (2001)].

\bibitem{Idilbi:2004vb}
  A.~Idilbi, X.~d.~Ji, J.~P.~Ma and F.~Yuan,
  Phys.\ Rev.\  D {\bf 70}, 074021 (2004).


\bibitem{Boer:2008fr}
  D.~Boer,
  Nucl.\ Phys.\  B {\bf 806}, 23 (2009).

\bibitem{Korchemsky:1994is}
  G.~P.~Korchemsky and G.~Sterman,
  Nucl.\ Phys.\  B {\bf 437}, 415 (1995).



\bibitem{Deshpande:2005wd}
  A.~Deshpande, R.~Milner, R.~Venugopalan and W.~Vogelsang,
  Ann.\ Rev.\ Nucl.\ Part.\ Sci.\  {\bf 55}, 165 (2005).

\bibitem{Adloff:1996dy}
 see, for example, C.~Adloff {\it et al.}  [H1 Collaboration],
  Nucl.\ Phys.\  B {\bf 485}, 3 (1997).



\bibitem{Kovchegov:1998bi}
  Y.~V.~Kovchegov and A.~H.~Mueller,
  Nucl.\ Phys.\  B {\bf 529}, 451 (1998).





\end{thebibliography}
\end{document}